\documentclass[twocolumn]{aastex62}

\shorttitle{Globular Cluster Systems of Coma UDGs}
\shortauthors{Lim et al.}

\begin{document}


\title{\bf \large The Globular Cluster Systems of Ultra-Diffuse Galaxies in the Coma Cluster}


\correspondingauthor{Sungsoon Lim, Eric Peng}
\email{slim@pku.edu.cn, peng@pku.edu.cn}

\author[0000-0002-5049-4390]{Sungsoon Lim}
\affiliation{Department of Astronomy, Peking University, Beijing 100871, China; slim@pku.edu.cn; peng@pku.edu.cn}
\affiliation{Kavli Institute for Astronomy and Astrophysics, Peking University, Beijing 100871, China}
\author[0000-0002-2073-2781]{Eric W. Peng}
\affiliation{Department of Astronomy, Peking University, Beijing 100871, China; slim@pku.edu.cn; peng@pku.edu.cn}
\affiliation{Kavli Institute for Astronomy and Astrophysics, Peking University, Beijing 100871, China}

\author[0000-0003-1184-8114]{Patrick C{\^o}t{\'e}}
\affiliation{Herzberg Astronomy \& Astrophysics Research Centre, National Research Council of Canada, Victoria, BC V9E 2E7, Canada}

\author[0000-0003-4073-7034]{Laura V. Sales}
\affiliation{Department of Physics and Astronomy, University of California, Riverside, CA 92521, USA}

\author{Mark den Brok}
\affiliation{Leibniz-Institute for Astrophysics (AIP), An der Sternwarte 16, 14482 Potsdam, Germany}

\author[0000-0002-5213-3548]{John P. Blakeslee}
\affiliation{Herzberg Astronomy \& Astrophysics Research Centre, National Research Council of Canada, Victoria, BC V9E 2E7, Canada}
\affiliation{Gemini Observatory, Casilla 603, La Serena 1700000, Chile}

\author[0000-0001-8867-4234]{Puragra Guhathakurta}
\affiliation{UCO/Lick Observatory, Department of Astronomy and Astrophysics, University of California Santa Cruz, 1156 High Street, Santa Cruz, CA 95064, USA}



\begin{abstract}
Ultra-diffuse galaxies (UDGs) are unusual galaxies with low luminosities, similar to classical dwarf galaxies, but sizes up to $\sim\!5$ larger than expected for their mass.
Some UDGs have large populations of globular clusters (GCs), something unexpected in galaxies with such low stellar density and mass. 
We have carried out a comprehensive study of GCs in both UDGs and classical dwarf galaxies at comparable stellar masses using HST observations of the Coma cluster. 
We present new imaging for 33 Dragonfly UDGs with the largest effective radii ($>2$ kpc), and additionally include 15 UDGs and 54 classical dwarf galaxies from the HST/ACS Coma Treasury Survey and the literature. Out of a total of 48 UDGs, 27 have
statistically significant GC systems, and 11 have candidate nuclear star clusters. 
The GC specific frequency ($S_N$) varies dramatically, with the mean $S_N$ being higher for UDGs than for classical dwarfs. At constant stellar mass, galaxies with larger sizes (or lower surface brightnesses) have higher $S_N$, with the trend being stronger at higher stellar mass. At lower stellar masses, UDGs tend to have higher $S_N$ when closer to the center of the cluster, i.e., in denser environments.  The fraction of UDGs with a nuclear star cluster also depends on environment, varying from $\sim\!40$\% in the cluster core, where it is slightly lower than the nucleation fraction of classical dwarfs, to $\lesssim20\%$ in the outskirts. Collectively, we observe an unmistakable diversity in the abundance of GCs, and this may point to multiple formation routes.

\end{abstract}





\section{Introduction}

Following the discovery of a population of faint, yet surprisingly large, galaxies in the Coma cluster \citep{vanD15}, ``ultra-diffuse galaxies" (UDGs, defined as galaxies with $\mu(g,0)\gtrsim24.0$ and $R_{\rm e,gal} \geq 1.5{\rm kpc}$) have been discovered in numerous environments. 
Although such galaxies have been known since the 1980s \citep{Bin85,Imp88} and a substantial number of similar types of galaxies have been previously identified \citep{Con03}, the survey of \citet{vanD15} has prompted an explosion of interest in these large, diffuse galaxies.
UDGs have been found in the other clusters \citep{Mih15,Mun15,vanB16,Lee17,Mih17,Jan17}, as well as in low-density environments \citep{Mar16,Mer16,Rom17,Tru17}. 
These galaxies have luminosities and morphologies similar to early-type dwarf galaxies, but larger sizes --- similar, in fact, to L* galaxies ($1.5<R_{\rm e,gal}<4.5$ kpc), making them outliers in the galaxy size-magnitude scaling relations. 

It remains unclear how UDGs formed and evolved. 
The very existence of UDGs in dense environments \citep{vanB16} suggests that at least some UDGs must be dominated by dark matter, and that environment may even play an important role in their evolution. UDGs may be ``failed" galaxies, meaning they were extremely inefficient at forming stars, and are thus severely under-luminous for their halo mass.
On the other hand, some suggest that UDGs are ``genuine" dwarf galaxies with correspondingly modest halo masses, but with anomalously large sizes. 
They could be the high spin tails of the halo angular momentum distribution \citep{Amo16}, or the result of feedback driven gas outflows that can lead to the expansion of both the dark matter and the stars \citep{DiC17}.

Globular clusters (GCs) are excellent probes of the stellar populations in these diffuse galaxies. 
GCs allow us to trace an early epoch of galaxy building when intense star formation was needed to form massive star clusters. 
Observationally, GCs are useful tracers of old stellar populations because they can be seen out to cosmological distances. 
Since the number of GCs in a galaxy correlates linearly with the total host halo mass \citep[e.g.,][]{Bla97,Pen08,Har17}, they provide a means of estimating the total mass of a galaxy based solely on photometric measures. While the sparse nature of UDGs points to inefficient star formation, GCs have, perhaps surprisingly, been identified in a number of UDGs.

Previous studies of GCs in UDGs have provided an important clue to their formation. 
\citet{Mih15} found a large GC population in VLSB-B, a UDG in the Virgo cluster. 
Its GC specific frequency ($S_N$)\footnote{Number of GCs per unit luminosity of host galaxy, $S_N=N_{GC}10^{0.4(M_V+15)}$} is considerably larger than those typical of dwarf galaxies with similar luminosity. 
Large GC populations have subsequently been reported for Virgo and Coma UDGs \citep{Bea16,Pen16,vanD16}, and in the field \citep{vanD18b}. These GCs may be metal-poor based on their broadband colors \citep{Pen06,Bea16b}. 
In addition, kinematic measurements of GCs indicate that at least some UDGs are dark matter dominated systems \citep{Bea16,vanD16,Tol18}, while others may have very little dark matter \citep{vanD18a}.
Taken together, these results show that UDGs are different from galaxies with similar luminosities. 

Nevertheless, UDGs may merely be large dwarf galaxies. 
There are several UDGs with highly elongated shapes, suggestive of tidal disruption \citep{Mer16,Mih17}. 
Equally important, a number of UDGs have no significant GC populations \citep{Mih15}. 

\citet{vanD17} and \citet{Amo18} have amassed fairly large UDG samples in the Coma cluster that are suitable for the study of GCs (15 and 18 UDGs, respectively) but these samples are nearly all in the cluster central regions (except for three UDGs from \citealt{vanD17}).  
Moreover, no previous GC study in UDGs has included a careful comparison with classical dwarf galaxies\footnote{We refer to more typical dwarf galaxies, those with low stellar mass and smaller sizes, as ``classical'' dwarf galaxies to avoid pre-supposing that UDGs are not dwarf galaxies} in the same environments, despite the fact that environment is known to affect GC specific frequency \citep{Pen08}. 
Thus, we need a comprehensive GC study of UDGs and classical dwarf galaxies located in the same environments. 
In this paper, we use the Hubble Space Telescope ({\it HST}) to analyze the GC systems for 33 Dragonfly UDGs that inhabit a range of environments within the Coma cluster. 
We also include twelve UDGs in the archival data of the Coma Treasury Survey \citep{Car08,Yag16}, as well as other measurements from the literature. 
For the comparison sample, we focus on classical dwarf galaxies selected from the Coma Treasury survey \citep{denB14}. 
In what follows, we adopt a distance to the Coma cluster of 100 Mpc ($m-M=35$; \citealp{Car08}), which we apply to all galaxies except for DF3, which is located behind the Coma cluster (142 Mpc, \citealp{Kad17}).

\section{Observations and Data Analysis}

\subsection{HST Imaging Program}
\begin{figure*}
\epsscale{1.2}
\plotone{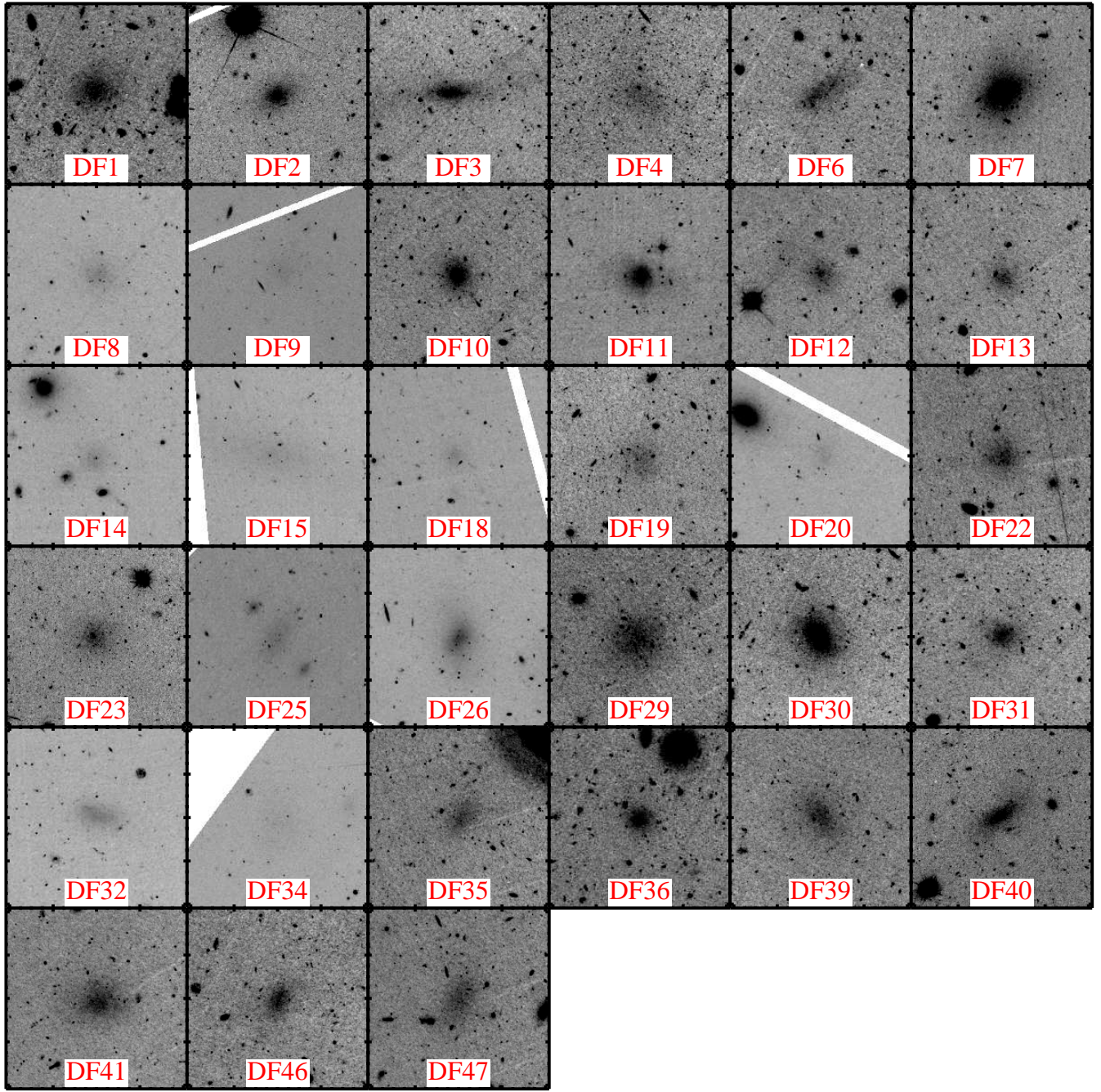}
\caption{Thumbnail images for our HST program sample. Each F606W image measures $50\arcsec\times50\arcsec$ and has been Gaussian smoothed with 5 pixel FWHM to show the UDG more clearly. North is up and east is to the left. \label{thumb}}
\end{figure*}

In a 26-orbit HST program (GO-14658), we imaged 33 large ($R_{\rm e,gal} \geq 2.0$~kpc) Coma cluster UDGs from the catalog of \citet{vanD15}. Each galaxy was observed for one orbit in the F606W (``wide $V$'') filter, with
ACS/WFC serving as the primary instrument, and WFC3/UVIS used in parallel mode. Accumulated exposure times were 2358 sec and 2445 sec for ACS/WFC and WFC3/UVIS, respectively. The observations were taken between January, 2017 and July, 2017.
CTE corrections and drizzling were performed with the standard STScI pipeline. 
Figure \ref{thumb} shows a montage of thumbnail images for our 33 program galaxies. 

Source catalogs were generated using the Source Extractor package \citep{Ber96}. 
Photometric magnitudes were obtained from aperture magnitudes measured within a 6 pixel ($0\farcs3$ for ACS, $0\farcs24$ for WFC3) diameter aperture. 
To correct for flux missed due to the finite aperture size, we computed aperture correction values out to a 10 pixel diameter ($0\farcs5$ for ACS, $0\farcs4$ for WFC3) using bright stars in the field, and then applied the aperture correction to infinity using values from the literature (\citealp{Boh11}; UVIS webpage\footnote{\url{http://www.stsci.edu/hst/wfc3/analysis/uvis_ee}}). 
We adopted photometric zeropoints for AB magnitudes in F606W for ACS/WFC and WFC3/UVIS from the appropriate STScI websites.

GCs in these galaxies appear as point sources because GCs at the distance of the Coma cluster are unresolved, even with HST resolution; we therefore identified point sources as candidate GCs. 
Similar to \citet{Pen11} and \citet{Pen16}, we adopted an ``inverse concentration" index ($C_{4-7}$), which is the difference in magnitude between a 4-pixel aperture ($0\farcs2$ for ACS, $0\farcs16$ for WFC3) and a 7-pixel aperture ($0\farcs35$ for ACS, $0\farcs28$ for WFC3) for the point source selection.
This parameter is normalized, so point sources will have $C_{4-7}=0$.
Most detected sources are brighter than $V_{606}\approx27.5$~mag, so we identified sources with $V_{606} \leq 27.5$~mag and $C_{4-7}<0.3$ as point sources.

We carried out artificial star tests to quantify the detection efficiency in our images. 
An empirical point spread function (PSF) was constructed using DAOPHOT II \citep{Ste87} using bright stars.
These empirical PSFs were added to the images as artificial stars; we then ran the same detection and selection procedures as were used to generate  the GC catalogs. 
The $90\%$ completeness levels were found to be $V_{606}=27.50$ mag and $V_{606}=27.55$ mag for ACS and WFC3, respectively. 

\subsection{UDGs in the Coma Cluster Treasury Survey}

The Coma Cluster Treasury Survey \citep{Car08} is a program that imaged the core of the Coma cluster in with HST in two filters, $g_{475}$ and $I_{814}$. 
The depth of the imaging is adequate to detect GCs at Coma distance down to the mean of the GC luminosity function \citep{Pen11}. These images contain 54 of the UDG candidates selected by \citet{Yag16} from ground-based imaging. This list, however, includes galaxies with $R_e$ as small as $0.7$~kpc, which is well within the range expected for normal dwarf galaxies. 

\begin{figure}
\epsscale{1.2}
\plotone{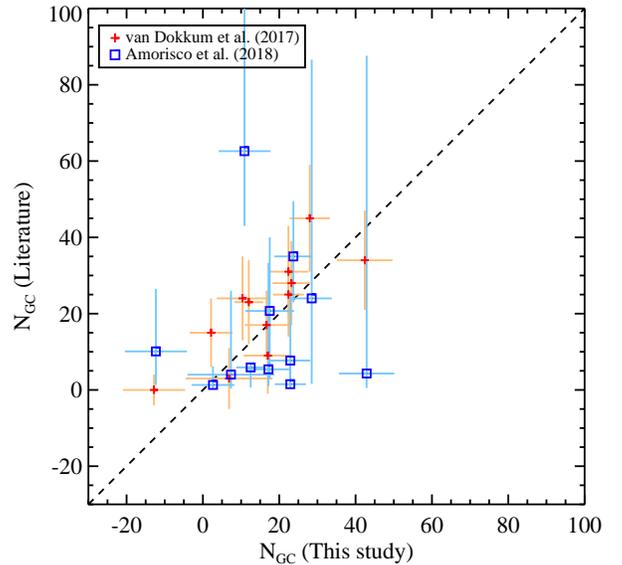}
\caption{GC numbers measured in this study compared with those measured for the same UDGs in two other published studies. The X-axis shows number of GCs in UDGs from this study, and Y-axis indicates number of GCs presented in the literature. Crosses and squares represent comparisons with \citet{vanD17} and \citet{Amo18}, respectively. Squares are shifted a little to avoid overlapping error bars. \label{comp}}
\end{figure}

\citet{vanD17} used the Coma Treasury Survey data, combining $g_{475}$- and $I_{814}$-band images to generate deeper pseudo-$V$ band images, and measured the sizes of these 54 galaxies using GALFIT \citep{Pen02}, finding that 12 of them have ${\rm R_{eff}}\geq1.5$ kpc. They then identified GC candidates associated with these 12 UDGs. 
\citet{Amo18} also studied GCs of UDGs in Coma Treasury Survey data. 
They provide estimates of the numbers of GCs in 54 UDG candidates from \citet{Yag16}, 18 of which have ${\rm R_{eff}}\geq1.5$ kpc based on the size measurement of \citet{Yag16}. 
Although these previous studies provide total GC numbers for UDGs in the Coma Treasury Survey data, we have re-measured them to ensure a consistent methodology across all galaxies.
We chose to include the 11 of the 12 UDGs from \citet{vanD17} in our analysis, excluding Y419 because it looks like superposition of two small galaxies (something mentioned in \citealt{vanD17}).
We also combined $g_{475}$- and $I_{814}$-band images to estimate the total number of GCs in these UDGs.
Our measured GC numbers are similar, within the uncertainties, to those from \citet{vanD17} and \citet{Amo18} (Figure \ref{comp}). 

In total, we have 45 UDGs with our consistent GC analysis after combining our new HST observations (33), UDGs from the Coma Treasury Survey (11), and DF17 from our previous work \citep{Pen16}. 
In what follows, we also included 3 UDGs (DF42, DF44, and DFX1) from \citet{vanD17} in our analysis, giving us a full sample of 48 UDGs.

\subsection{A Comparison Sample of Classical Dwarf Galaxies}
To understand the properties and origins of these UDGs, a clean comparison with classical dwarf galaxies is important. 
There are numerous studies of globular clusters in classical dwarf galaxies, but no large samples for the Coma cluster. 
We therefore performed the same GC analysis for classical dwarf galaxies in the Coma Treasury Survey using the catalog of \citet{denB14}. 

\citet{denB14} studied galaxy photometric parameters, including magnitudes and effective radii, using the Coma Treasury Survey data. 
Their study contains about 200 galaxies within the magnitude range of $-18 \lesssim M_V \lesssim -12$. We used 88 of these galaxies in our GC analysis, after excluding galaxies located in the halos of bright galaxies or close to the edges of the HST fields. Of these 88 galaxies, 54 have magnitudes in the range spanned by the UDGs, and we use these to provide a matched comparison sample for the UDGs. 

\subsection{Globular Cluster Numbers and Specific Frequencies}
Measuring the total number of GCs in a galaxy requires us to make an estimate (or assumption) about the spatial extent and luminosity function of the GCs. In \citet{Pen16}, we estimated the spatial extent of GCs in DF17 directly from the data. Unfortunately, it is difficult to measure GC system spatial extent in these UDGs due to the comparatively shallower imaging. 
Although \citet{For17} fit a relation between $R_{\rm e,GCS}$ and host galaxy stellar mass for massive early-type galaxies, he also noted that UDGs do not follow the relation.
\citet{vanD17} adopted a relation between the effective radii of the galaxy and GC system, concluding that $R_{\rm e,GCS} \sim 1.5R_{\rm e,gal}$ is a reasonable approximation for UDGs where both quantities are measured. 
We have tested this assumption using GC and galaxy data from the ACS Virgo Cluster Survey \citep[ACSVCS;][]{Cot04,Fer06,Jor09}, and find it to be an adequate description for low-mass galaxies, so we have used the same assumption to derive total GC numbers in our analysis. 

The magnitude limit of our data is similar to the peak magnitude of the GC luminosity function (GCLF) at the distance of the Coma cluster, so it is difficult to constrain the shape or mean of the GCLF. 
Therefore, we have assumed that the shape of GC luminosity function in UDGs is the same as that in classical dwarf galaxies \citep[i.e., roughly Gaussian;][]{Jor07,Mil07}. In DF17, where we could measure the GCLF past the turnover, we found that this was a reasonable assumption \citep{Pen16}.

We measure the number of GC candidates with $g<27.5$~mag within $1.5R_{\rm e,gal}$. This number will be contaminated by compact background sources and intra-cluster GCs (IGCs). 
We estimate the level of this contamination using a local background in an annulus outside $5R_{\rm e,gal}$. 
After background subtraction and completeness corrections down to the GCLF peak, we obtain the total number of GCs within the aperture by multiplying by two, which is the traditional way to account for the unseen faint end of the GCLF in extragalactic systems \citep{Har81}. We then multiple by two again to account for the GCs that are outside our chosen aperture of $1.5R_{\rm e,gal}$.

To check the reliability of the latter assumptions, 
we also estimated the total number of GCs by measuring the number of GCs within $4R_{\rm e,gal}$. 
Based on the assumption that $R_{\rm e,GCS} \sim 1.5R_{\rm e,gal}$, the fraction of GCs within $4R_{\rm e,gal}$ is about $94\%$. 
Thus, we calculated the total number of GCs based on this measurement, comparing the results to that obtained
from GC counts within $1.5R_{\rm e,gal}$ and $4R_{\rm e,gal}$. 
The two estimates are in good agreement with low scatter, so we conclude that the assumption $R_{\rm e,GCS} \sim 1.5R_{\rm e,gal}$ is reliable for UDGs.

For classical dwarf galaxies, we used the same assumption for the GC distribution, but estimated the total number of GCs within $4R_{\rm e,gal}$. Because the effective radii of classical dwarf galaxies are generally small, this smaller size allow us to use a larger aperture (relative to $R_{e,gal}$) without incurring a large penalty from the background. 
After applying corrections for completeness and sampling of the GC luminosity function, we obtain the total number of GCs by multiplying by a factor of 1.06 to correct GC fraction within $4R_{\rm e,gal}$ to total number.

In all galaxies, we estimated the amplitude of fluctuations in background counts. 
To estimate this fluctuation, we randomly chose locations on the HST images excluding regions with UDGs and Coma galaxies \citep{Soh17}, and measured the number of GC candidates using same size aperture as for the background annulus of the target UDG. 
We repeated this procedure for more than a hundred locations and used the standard deviation of these values as the background fluctuation amplitude.  
The final uncertainties in the total GC numbers are a combination of Poisson errors and the estimated background fluctuations.

\begin{figure}
\epsscale{1.2}
\plotone{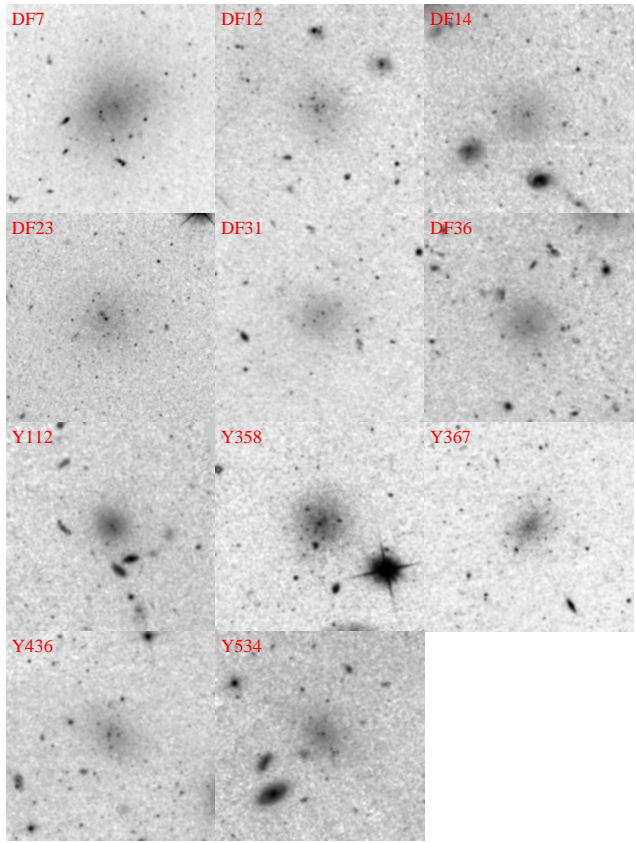}
\caption{Thumbnail images for eleven nucleated UDGs. To highlight subtle morphological details, we show $30\arcsec \times 30\arcsec$ regions centered on each UDG. Images have been smoothed with a Gaussian kernel of $FWHM=7$~pixels to emphasize faint structures. UDG names are given at the upper left corner of each panel. North is up and east is to the left. \label{nuc}}
\end{figure}

Finally, we note that we find eleven UDGs that may contain central nuclei (Figure \ref{nuc}) i.e.,
UDGs with bright point-like sources at their photocenters. 
The candidate nuclei in DF7 and DF12 are not selected as GCs, but those in the remaining galaxies were identified as GCs by our automated selection. In all cases, we excluded these nuclei when calculating the number of GCs.
The GC system properties of the Coma UDGs and classical dwarf galaxies are summarized in Tables \ref{tbl:udg} and \ref{tbl:dwarf}, respectively.

\section{Results}

\begin{figure}
\epsscale{1.25}
\plotone{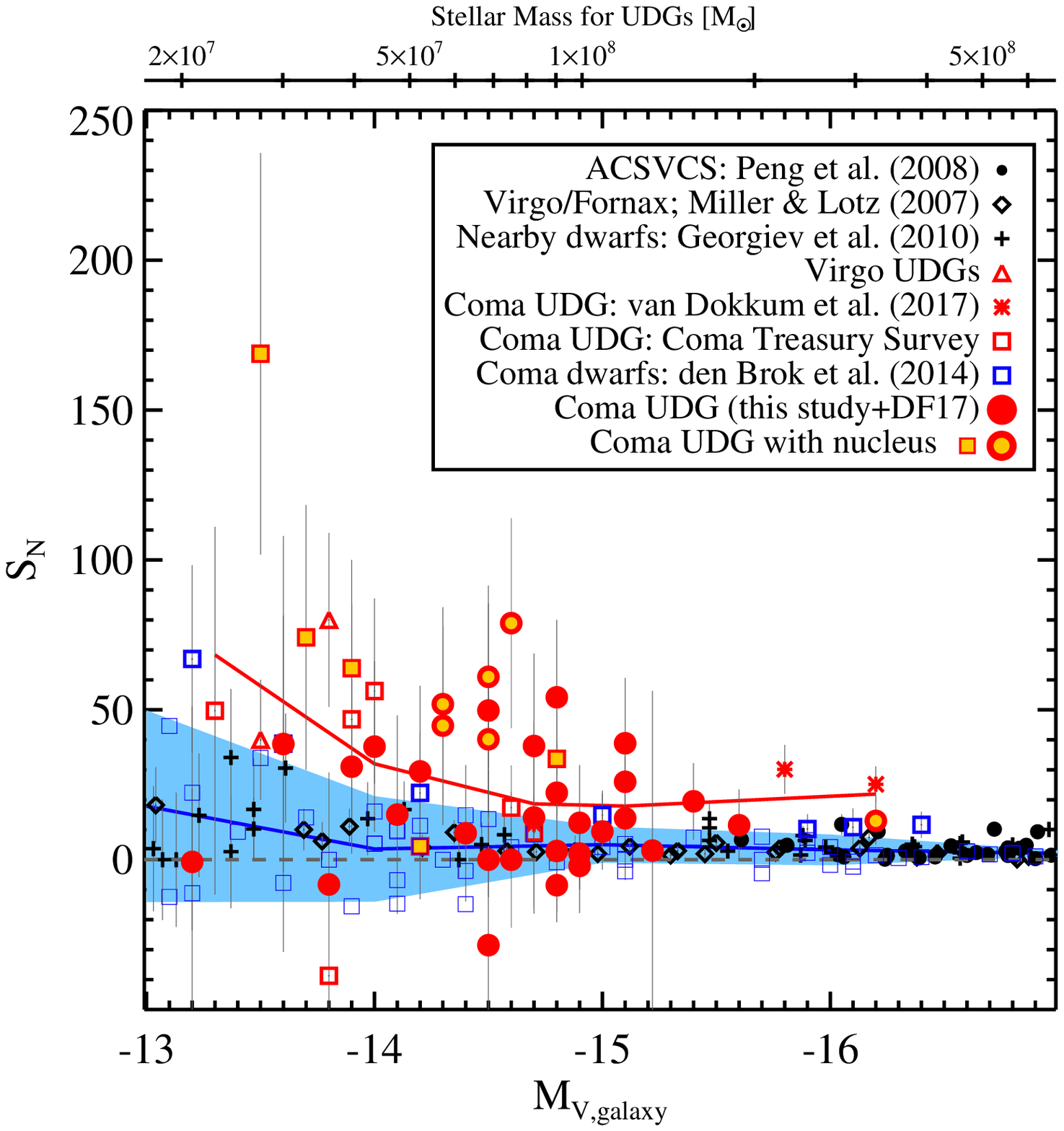}
\caption{Globular cluster specific frequency, $S_N$, versus galaxy $V$-band absolute magnitude. Plotted are Virgo cluster galaxies from the ACSVCS \citep[][black dots]{Pen08}, dwarf galaxies in the Virgo and Fornax clusters \cite[][black diamonds]{Mil07}, and nearby dwarf galaxies \citep[][crosses]{Geo10}. Literature UDGs are plotted for Virgo \citep[][red triangles]{Mih15} and Coma \citep[][red asterisks]{vanD17}. Red filled circles and red open squares show UDGs from our HST program and the Coma Treasury Survey data, respectively. We have indicated nucleated UDGs as the small yellow filled circles. The red solid line shows the mean $S_N$ values for all Coma UDGs. The blue solid line shows the mean $S_N$ values for ``classical'' Coma dwarf galaxies from \citet{denB14} (shown individually as blue squares), and blue shading shows their associated standard deviation ($1\sigma$). Most UDGs (i.e., all red symbols) have large $S_N$ values that fall above the $1\sigma$ distribution for classical Coma dwarf galaxies. The grey dashed line indicates $S_N=0$. Along the top axis, we show the stellar masses for the UDGs calculated from their $V-$band total magnitudes using \citet{Bel01}, assuming their colors to be the same as DF44 and DFX1.  \label{mvsn}}
\end{figure}

Figure \ref{mvsn} compares $S_N$ for UDGs and classical dwarf galaxies, plotting their specific frequencies against absolute $V$-band magnitudes. 
On average, the UDGs have higher $S_N$ values than the classical dwarfs. 
At a given luminosity or stellar mass, the mean $S_N$ values of UDGs are more than 1-sigma above the mean $S_N$ values of classical dwarf galaxies. 
The upper range of $S_N$ values for UDGs are from $\sim\!30$ to $\sim\!100$, for $-16<M_V<-13$~mag. 
These values are extremely high compared to any other type of galaxy, suggesting that UDGs have some of the largest GC fractions among known galaxies. 
At the same time, several UDGs have $S_N$ values similar to those of classical dwarfs, and some others are consistent with having no GCs at all. UDGs clearly exhibit a very wide range in GC properties.
Our focus in the next section is to understand the origin this diversity.

\subsection{GC Specific Frequencies and Connections to Host Galaxy Properties}

\begin{figure}
\epsscale{1.25}
\plotone{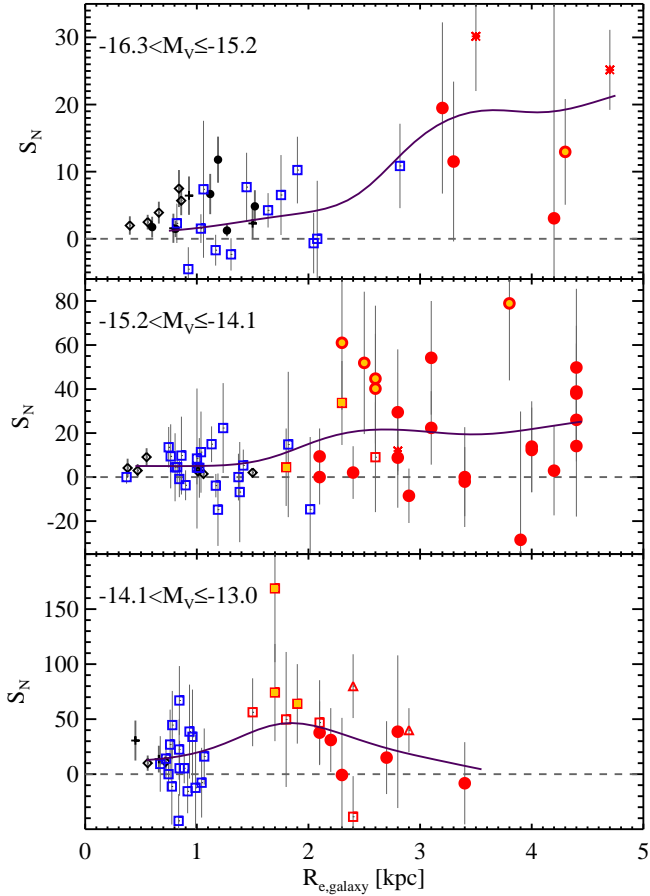}
\caption{$S_N$ plotted as a function of galaxy effective radius. All symbols are the same as in Figure \ref{mvsn}. We divided galaxies into three groups according to their total magnitudes, and magnitude ranges are noted in the top left corners of panels. Purple solid lines show mean $S_N$ values of all Coma galaxies in each magnitude range. The mean lines are smoothed using Gaussian kernel with FWHM of 0.5 kpc. Grey dashed lines indicate $S_N =0$.  
\label{resn}}
\end{figure}

We have examined the properties of the individual galaxies in order to better understand which parameters are responsible for the observed wide range in specific frequency. 
Although UDGs have large sizes, their luminosities are similar to classical dwarf galaxies, leading us to suspect
that the high specific frequencies are somehow related to the larger sizes of UDGs relative to classical dwarfs.
Figure \ref{resn} shows our measured $S_N$ values plotted as a function of effective radius in three bins of galaxy magnitude. 
We find that for the more luminous galaxies, mean $S_N$ increases with galaxy size, although there is large scatter. 
There is no corresponding trend for the fainter galaxies, however.
In order to make sure that our size-dependent aperture for counting the number of GCs was not causing the trend for the brighter galaxies, we also used a fixed large aperture to count the number of GCs and we found the same trend.

\begin{figure}
\epsscale{1.25}
\plotone{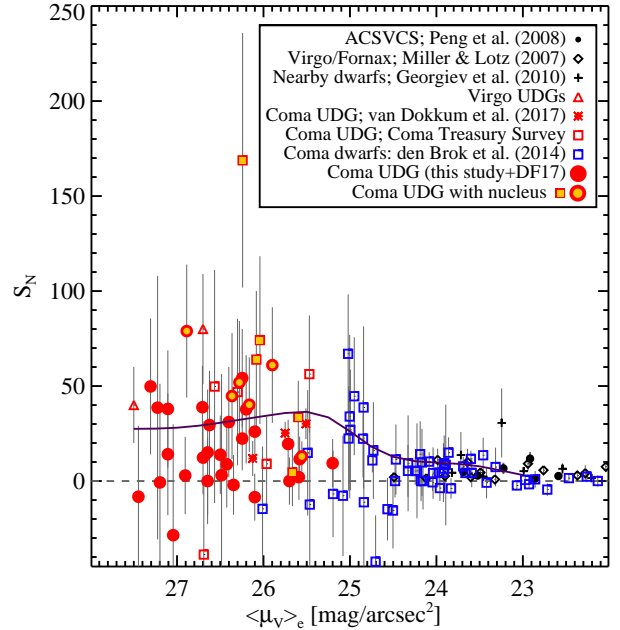}
\caption{$S_N$ plotted as a function of galaxy Surface brightness.  All symbols are the same as in Figure \ref{mvsn}. The purple solid line shows mean values of $S_N$ for all Coma galaxies (i.e., UDGs plus classical dwarfs). The line for mean values are smoothed using Gaussian kernal with FWHM of 0.5 mag$/$arcsec$^2$. The grey dashed line indicates $S_N=0$. There is no significant correlation between surface brightness and $S_N$. \label{musn}}
\end{figure}

Surface brightness is another extreme property of UDGs, so we 
examine the relationship between $S_N$ and surface brightness in Figure \ref{musn}. 
There appears to be a transition, where galaxies with surface brightness is fainter than $\langle\mu_V\rangle_e \sim 25$~mag arcsec$^{-2}$ have higher $S_N$ values than those in higher surface brightness galaxies, although there is a lot of scatter in this trend as well. 

\subsection{GC Specific Frequencies and their Connection to Environment}

\begin{figure}
\epsscale{1.25}
\plotone{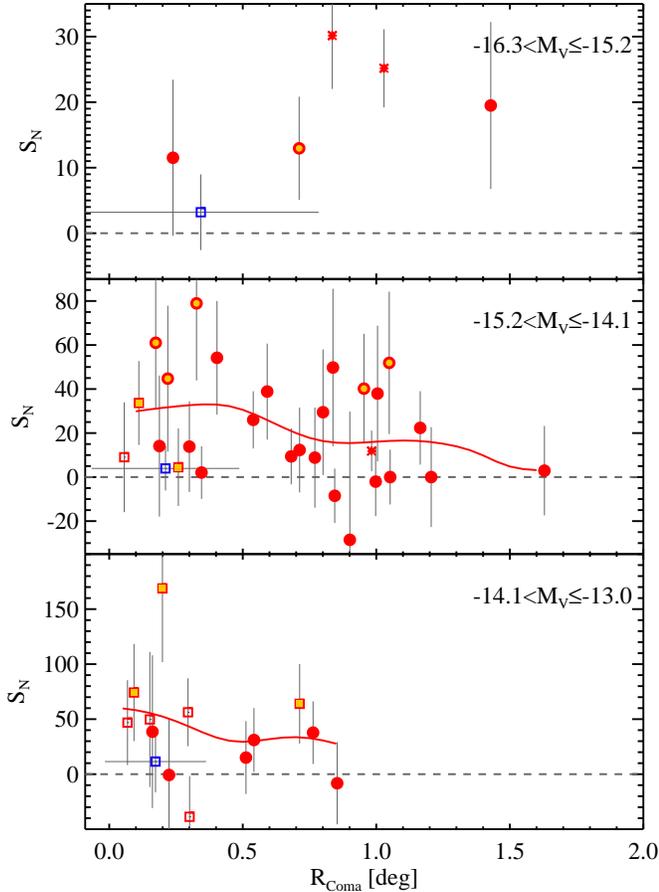}
\caption{$S_N$ of Coma galaxies plotted as against distance from cluster center (taken to be NGC 4874). All symbols are the same as in Figure~\ref{mvsn}, but we only plotted UDGs in the Coma cluster and mean value of classical Coma dwarf galaxies (blue squares, RMS are shown as error bars) in this figure. The UDGs are divided into three groups with their total magnitude as same as Figure \ref{resn}. Solid lines show mean $S_N$ values, and they are smoothed using Gaussian kernel with FWHM of 0.15 deg.  
\label{rcomasn}}
\end{figure}

There is also evidence that environment can also affect GC specific frequencies, particularly for low mass galaxies. In Virgo cluster dwarfs, $S_N$ is generally higher the closer the galaxy is to the cluster center \citep{Pen08}. 
Figure \ref{rcomasn} shows $S_N$ plotted against distance from the Coma cluster center, ${\rm R_{Coma}}$, which we take to be the cD galaxy, NGC 4874. As in Figure~\ref{resn}, we divide the sample into three bins of $M_V$ in order to separate out the trends in magnitude that we see in Figure~\ref{mvsn}.

The $S_N$ of the most luminous UDGs do not show any trend with $R_{Coma}$, but with only five galaxies in this bin it is difficult to draw any conclusions. The fainter bins, however, do appear to display a correlation between $S_N$ and proximity to the cluster center. In the intermediate luminosity sample, the {\it mean} $S_N$ in the cluster core is $\sim\!30$, and in the faintest bin, the mean $S_N$ is $\sim\!60$. These are very high $S_N$ values when compared to the classical dwarfs in this luminosity range. In both magnitude bins, there is a weak but noticeable trend, with large scatter, for the mean $S_N$ to be higher at the cluster center and lower in the outskirts.

\subsection{Nuclear star clusters}
\begin{figure}
\epsscale{1.25}
\plotone{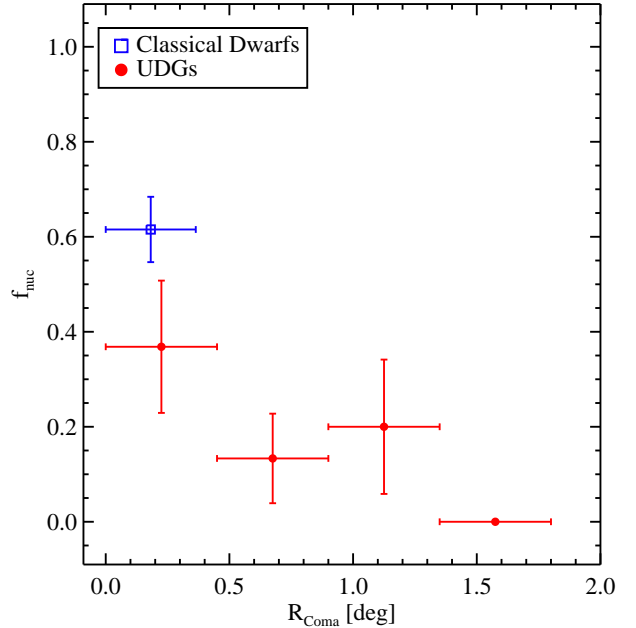}
\caption{Fraction of nucleated UDGs and classical dwarfs plotted against distance from cluster center. Red filled circles indicate nucleated UDG fraction, and blue open square shows nucleation fraction of classical dwarf galaxies within our UDG magnitude range from \citet{denB14}. Error bars on X-axis show bin sizes, and error bars on Y-axis indicate pure Poisson errors. \label{nfrac}}
\end{figure}

\citet{denB14} studied the nucleation fraction of classical dwarf galaxies in the Coma cluster core and showed that is was quite high, with $f_{nuc}\approx60\%$. We visually identified stellar nuclei in our UDG sample, and found that 11 out of 48 galaxies appeared nucleated, an overall nucleation fraction of $f_{nuc,UDG}=23\%$. At first, this seems significantly lower than for the classical dwarfs, but there has been some evidence that $f_{nuc}$ may be higher in denser environments \citep{Bal14,Ord18}, and the Coma classical dwarf sample is biased toward the cluster core. In Figure~\ref{nfrac}, we plot the UDG nucleation fraction as a function of $R_{Coma}$ and find a distinct trend where the $f_{nuc,UDG}\approx 40\%$ in the cluster core, dropping to 0--20\% in the outskirts. Even at the center of the cluster, UDGs still seem to have a lower nucleation fraction than the classical dwarfs, although the difference is not as large as when comparing with the full UDG sample.

\begin{figure}
\epsscale{1.25}
\plotone{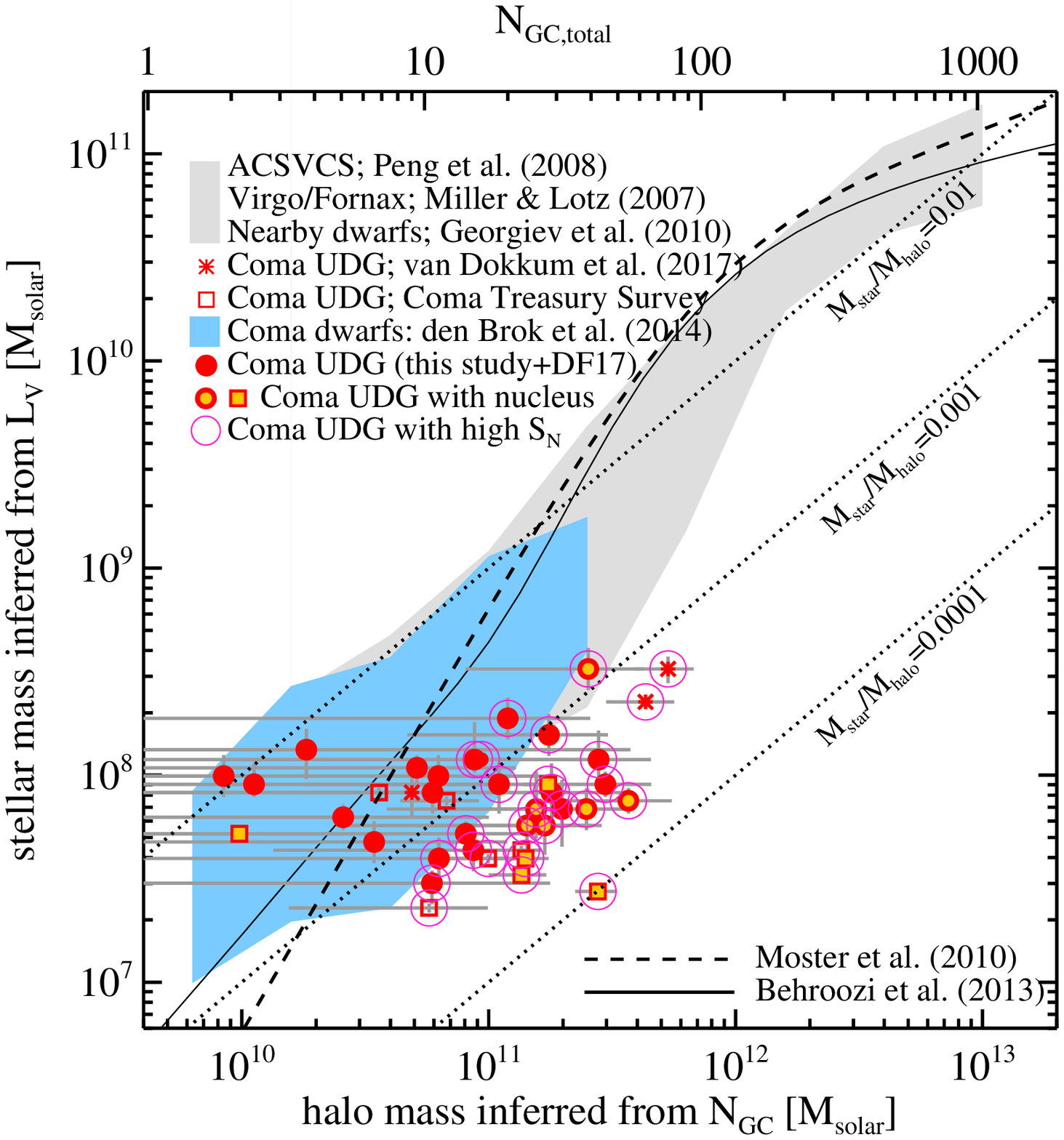}
\caption{Inferred stellar mass vs.\ inferred total mass. The ordinate shows the stellar mass measured from the total $V-$band magnitudes using \citet{Bel01}, while the abscissa is total mass inferred from the $N_{GC}$--total mass relation of \citet{Har17}. We also note the total number of GCs along the upper axis. Filled circles show UDGs from this study and \citet{Pen16}. Open squares and asterisks show UDGs from the Coma Treasury Survey and \citet{vanD17}, respectively. Nucleated UDGs are indicated by small yellow dots, and UDGs with high $S_N$ values are highlighted with large open circles. Grey shaded regions shows galaxies from other environments. The blue shaded region indicates classical dwarf galaxies in the Coma cluster. The dashed line and solid line show the expectation from abundance matching and their extrapolation below $M_{\star} \sim10^8 M_{\odot}$ \citep{Mos10} and $M_{halo}\sim10^{10}M_{\odot}$ \citep{Beh13} \label{tmass}}
\end{figure}

\begin{figure}
\epsscale{1.2}
\plotone{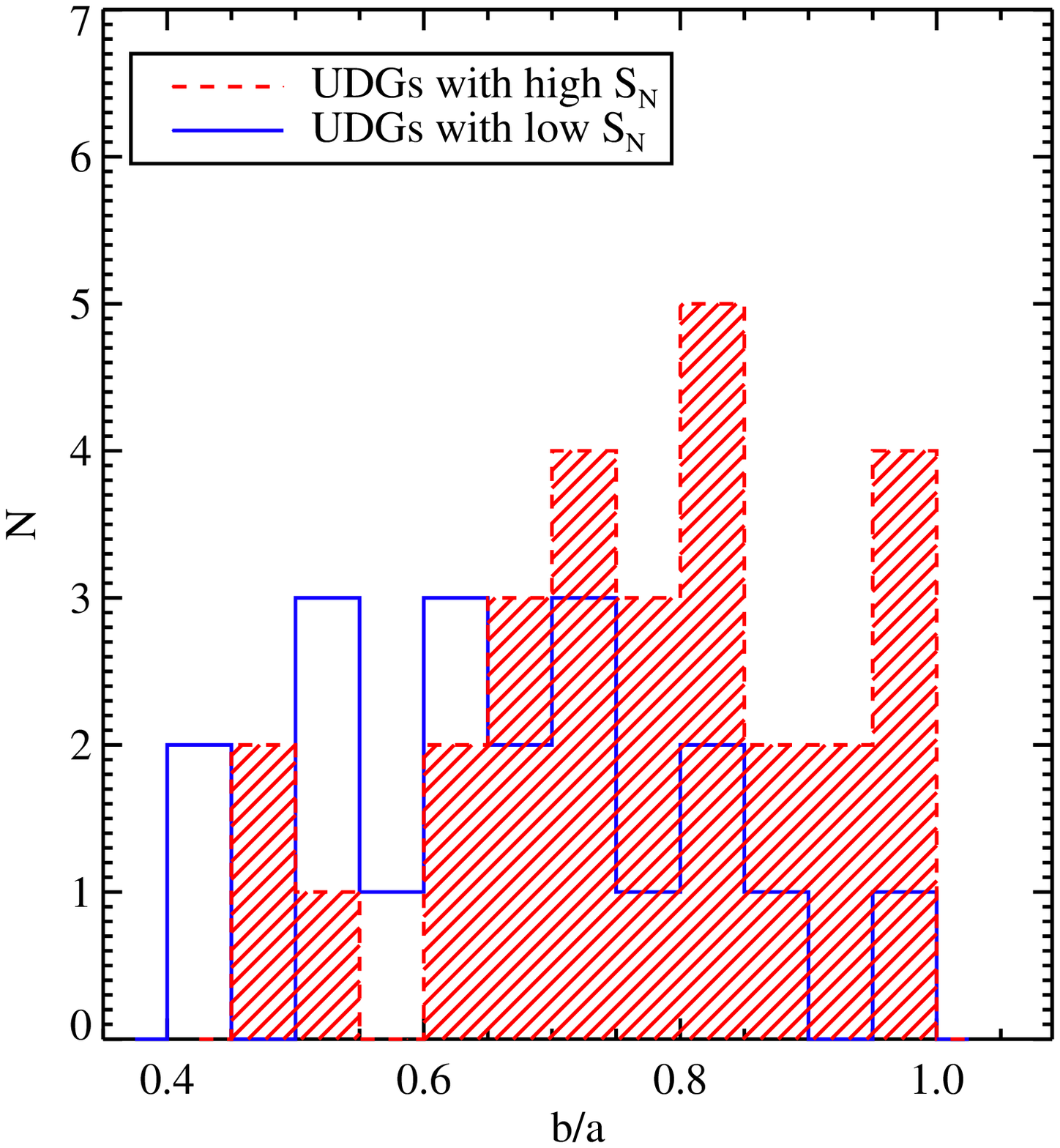}
\caption{Distribution of axial ratios, $b/a$, for UDGs. The $b/a$ values are from \citet{vanD15} and \citet{vanD17}. The blue solid open histogram shows the distribution for UDGs with $S_N$ values similar to classical dwarf galaxies. The red dashed histogram shows the distribution for UDGs with high $S_N$ values. A KS test for these two distributions yields a K-S statistic of 0.40 and probability of 0.038, indicating that the distributions differ at high confidence. \label{baudg}}
\end{figure}

\section{Discussion}

We have shown that UDGs exhibit a wide range in GC content, and that they generally have more GCs per unit luminosity than their classical dwarf counterparts. Some UDGs have $S_N$ values well above those of most classical dwarf galaxies; at the same time, nearly half of UDG sample shows $S_N$ values that are indistinguishable from those in classical dwarfs. 
We divide UDGs into ``low" $S_N$ UDGs --- with values within $1\sigma$ of that in classical dwarf galaxies (e.g., the blue region in Figure~\ref{mvsn}--- and ``high" $S_N$ UDGs --- with values which are higher than $1\sigma$ limits observed in classical dwarf galaxies. 
Interestingly, most nucleated UDGs (10/11) are members of the high $S_N$ class. 

We can also investigate the total masses of UDGs using their GC properties. 
Several studies have shown that the number of GCs is related to the total mass of a galaxy \citep{Bla97,Pen08,Har17}, and UDGs may obey this relation based on dynamical mass estimates for three UDGs \citep{Bea16,vanD17}. 
We cannot overstate that the number of GCs is a highly uncertain measure of galaxy halo mass, especially in this mass regime, but it is the only information we have for a vast majority of UDGs. We have transformed the number of GCs to halo masses using the relation, ${\rm log} M_{\rm halo} = 9.62+1.12 {\rm log} N_{\rm GC}$ \citep{Har17}. 
Although the interpretation of halo mass for satellites is not straightforward, it is customary to consider it as an estimate of the virial mass of their halos possibly at the time of infall into the cluster. With that caveat in mind, we find that the median total mass is about $10^{11} M_{\odot}$, which is larger than expected at their luminosity compared to simple extrapolation of abundance matching results (e.g. \citealp{Mos10,Beh13}), but still within the dwarf halo regime.

Figure \ref{tmass} shows the relation between stellar mass and total mass. The stellar masses were measured from the galaxy luminosities and colors using \citet{Bel01}. There are many UDGs offset to higher halo mass when compared to the region defined by normal galaxies (blue and gray shaded regions) . The UDGs have a median inferred $M_h/L_V \sim1000$, suggestive of a massive galaxy origin. These galaxies could have formed their GCs early, after which time star formation was quenched preventing the formation of a disk and bulge. As our UDGs are located in a rich galaxy cluster environment, one might suppose that UDGs were efficiently quenched once they fell into the cluster potential well. A similar process can explain variation in $S_N$ for classical dwarfs \citep{Pen08,Mis16,Liu16}.

The average total mass of high-$S_N$ UDGs is similar to that of M33. 
The $R_{\rm e,GCS}$ of the M33 GC system ($\sim\!3\ {\rm kpc}$, \citealp{Pen16}) is also within the range of the $R_{\rm e,GCS}$ used for UDGs in this study (median $R_{\rm e,GCS}\sim\! 4\ {\rm kpc}$). 
As a thought experiment, we can crudely estimate the infall time of these UDGs using M33's star formation history. 
UDGs have about $5\%$ of the stellar mass of M33, and the time when M33 formed $5\%$ of current stellar mass is about $7$~Gyr ago based on its inner halo region of \citet{Bar11}. Being in a denser environment, it is likely that the Coma UDGs built their stars up faster. It is then plausible that galaxies having a mass similar to M33 fell into the Coma cluster at least 7~Gyr ago ($z_{infall}>0.8$), and that their star formation was subsequently quenched by environmental processes, meaning that these galaxies could not form additional stars and eventually became UDGs.
This qualitative picture is also supported by recent spectroscopic studies of UDGs in our sample which find mean stellar ages of 7--9 Gyr \citep{Gu17,Fer18,Rui18}.

Low-$S_N$ UDGs, on the other hand, lie on the expected relation between stellar mass and total mass in Figure \ref{tmass}. These systems could have halo masses similar to those of classical dwarf galaxies. 
Indeed, a dwarf galaxy origin scenario for UDGs has been raised in several studies \citep[e.g.,][]{Amo16,DiC17,Ron17}. 
Total inferred halo masses for low-$S_N$ UDGs have a similar range as halo masses for UDGs with a dwarf galaxy origin seen in simulations. 
These UDGs could represent the high angular momentum tail of dwarf galaxies; alternatively, feedback driven gas outflows or tidal harassment, could expand both the stellar content and dark matter within dwarf galaxies.

Finally, we show in Figure~\ref{baudg} that low-$S_N$ UDGs are on average slightly more elongated than high-$S_N$ UDGs. The distributions of $b/a$ for high- and low $S_N$ UDGs are moderately different (at $\gtrsim2\sigma$), with a KS test indicating $D=0.40$ and probability of $p_{KS}=0.038$.  This could be an indication having higher angular momentum, but this is speculation until we have stellar kinematic observations. 

\section{Summary and Conclusion}
We have presented a comprehensive study of GCs in 48 UDGs in the Coma cluster using HST data. Our sample includes 33 UDGs from a new observing program, one from our previous HST study \citep{Pen16}, 11 from the Coma Treasury Survey, and three UDGs from \citet{vanD17}. Our primary findings can be summarized as follows. 

\begin{enumerate}
\item The GC specific frequencies of UDGs are found to vary dramatically. On the whole, though, the mean $S_N$ values of UDGs are higher than those of classical dwarf galaxies at a given luminosity.

\item For the most luminous subset of UDGs, we find that GC specific frequencies are higher when the galaxy effective radius is larger. 
This trend is not so apparent for fainter UDGs.

\item For intermediate- and low-luminosity UDGs, those closer to the cluster center have higher $S_N$ values, pointing to the importance of environmental processes. 

\item We have divided UDGs into two classes: UDGs with high and low $S_N$ values as compared to those of classical dwarf galaxies. The subsample of high-$S_N$ UDGs are dark matter dominated systems with $M/L_V \gtrsim 1000$ based on the total mass inferred from total number of GCs. UDGs with low GC specific frequencies have $M/L_V$ values similar to those of classical dwarf galaxies ($M/L_V\sim500$). 

\item We find eleven of our 48 UDGs to harbor nuclear star clusters. The fraction of nucleated UDGs varies with the distance to center of the cluster. UDGs have a higher nucleation fraction ($f_{nuc,UDG}\approx40\%$) in the cluster core, decreasing to 0--20\% in the outskirts. Classical dwarf galaxies, however, have an even higher nucleation fraction, $f_{nuc}\approx 60\%$ \citep{denB14}, in the cluster core. Ten of the eleven nucleated UDG candidates are classified as high-$S_N$ systems.

\end{enumerate}

The diversity in GC systems suggests that UDGs may arise through more than one formation channel \citep{Pan17,Amo18,Ala18}.
Further study of UDGs, particularly kinematic constraints from deep IFU spectroscopy, could provide important clues for our understanding of the properties and origins of UDGs. 

\acknowledgments
SL thanks Jubee Sohn for providing a galaxy catalog for the Coma cluster. 
SL and EWP acknowledge support from the National Natural Science Foundation of China through Grant No.\ 11573002, and from the Strategic Priority Research Program, ``The Emergence of Cosmological Structures,'' of the Chinese Academy of Sciences, Grant No. XDB09000105. 
LVS acknowledge support from the Hellman Foundation and HST grant HST-AR-14583. 
Based on observations made with the NASA/ESA Hubble Space Telescope, obtained [from the Data Archive] at the Space Telescope Science Institute, which is operated by the Association of Universities for Research in Astronomy, Inc., under NASA contract NAS 5-26555. These observations are associated with program 14658.

\facility{HST}

\begin{table*}
\small
\begin{center}
\caption{The globular cluster system properties of Coma ultra-diffuse galaxies\label{tbl:udg}}
\begin{tabular}{lcccccc}
\tableline\tableline Name & $N_{\rm GC}$ & $S_N$ & $M_V$ & $R_{\rm e,gal}$ & $\langle\mu_V\rangle_e$ & $R_{\rm Coma}$  \\
   &  &  & [mag] & [kpc] & [mag/arcsec$^2$] & [deg] \\
\tableline
 DF1& $ 18.6 \pm  13.9$ & $ 22.3 \pm  16.7$ & $-14.8 $ &  3.1 &  26.2 &  1.164 \\
 DF2& $  0.0 \pm   7.8$ & $  0.0 \pm  12.4$ & $-14.5 $ &  2.1 &  25.7 &  1.052 \\
 DF3$^a$&$  3.7 \pm  65.2$ & $  3.1 \pm  53.2$ & $-15.2 $ &  4.2 &  26.5 & \nodata \\
 DF4& $-18.0 \pm  36.8$ & $-28.5 \pm  58.3$ & $-14.5 $ &  3.9 &  27.0 &  0.901 \\
 DF6& $ 31.4 \pm  22.6$ & $ 49.8 \pm  35.8$ & $-14.5 $ &  4.4 &  27.3 &  0.838 \\
 DF7$^b$& $ 39.1 \pm  23.8$ & $ 12.9 \pm   7.9$ & $-16.2 $ &  4.3 &  25.6 &  0.711 \\
 DF8& $ 42.6 \pm  23.9$ & $ 38.9 \pm  21.8$ & $-15.1 $ &  4.4 &  26.7 &  0.591 \\
 DF9& $ 14.1 \pm  13.7$ & $ 29.5 \pm  28.5$ & $-14.2 $ &  2.8 &  26.6 &  0.800 \\
DF10& $  1.9 \pm  10.9$ & $  2.1 \pm  12.0$ & $-14.9 $ &  2.4 &  25.6 &  0.346 \\
DF11& $  9.4 \pm  12.7$ & $  9.4 \pm  12.7$ & $-15.0 $ &  2.1 &  25.2 &  0.681 \\
DF12$^b$& $ 23.5 \pm  17.4$ & $ 44.7 \pm  33.1$ & $-14.3 $ &  2.6 &  26.4 &  0.219 \\
DF13& $ 11.3 \pm  10.5$ & $ 31.0 \pm  29.0$ & $-13.9 $ &  2.2 &  26.4 &  0.542 \\
DF14$^b$& $ 54.6 \pm  24.2$ & $ 78.9 \pm  35.0$ & $-14.6 $ &  3.8 &  26.9 &  0.327 \\
DF15& $ 15.2 \pm  22.6$ & $ 13.8 \pm  20.6$ & $-15.1 $ &  4.0 &  26.5 &  0.300 \\
DF18& $ 10.6 \pm  19.1$ & $ 38.6 \pm  69.4$ & $-13.6 $ &  2.8 &  27.2 &  0.162 \\
DF19& $ 28.8 \pm  23.4$ & $ 37.9 \pm  30.8$ & $-14.7 $ &  4.4 &  27.1 &  1.005 \\
DF20& $ -0.1 \pm   9.9$ & $ -0.7 \pm  51.9$ & $-13.2 $ &  2.3 &  27.2 &  0.224 \\
DF22& $ 15.0 \pm  11.3$ & $ 37.7 \pm  28.5$ & $-14.0 $ &  2.1 &  26.2 &  0.763 \\
DF23$^b$& $ 38.5 \pm  19.2$ & $ 61.0 \pm  30.4$ & $-14.5 $ &  2.3 &  25.9 &  0.174 \\
DF25& $ 10.7 \pm  24.3$ & $ 14.1 \pm  32.0$ & $-14.7 $ &  4.4 &  27.1 &  0.188 \\
DF26& $ 20.0 \pm  20.7$ & $ 11.5 \pm  11.9$ & $-15.6 $ &  3.3 &  25.6 &  0.239 \\
DF29& $ 45.1 \pm  21.5$ & $ 54.2 \pm  25.8$ & $-14.8 $ &  3.1 &  26.2 &  0.404 \\
DF30& $ 28.2 \pm  18.4$ & $ 19.5 \pm  12.8$ & $-15.4 $ &  3.2 &  25.7 &  1.429 \\
DF31$^b$& $ 27.2 \pm  17.0$ & $ 51.9 \pm  32.3$ & $-14.3 $ &  2.5 &  26.3 &  1.048 \\
DF32& $  5.1 \pm  13.1$ & $  8.8 \pm  22.8$ & $-14.4 $ &  2.8 &  26.4 &  0.770 \\
DF34& $ -2.7 \pm  12.4$ & $ -8.3 \pm  37.4$ & $-13.8 $ &  3.4 &  27.4 &  0.854 \\
DF35& $  6.6 \pm  14.4$ & $ 15.1 \pm  33.1$ & $-14.1 $ &  2.7 &  26.6 &  0.512 \\
DF36$^b$& $ 25.4 \pm  15.7$ & $ 40.2 \pm  24.9$ & $-14.5 $ &  2.6 &  26.2 &  0.954 \\
DF39& $ 11.2 \pm  17.6$ & $ 12.3 \pm  19.3$ & $-14.9 $ &  4.0 &  26.7 &  0.713 \\
DF40& $ -7.1 \pm  10.3$ & $ -8.5 \pm  12.4$ & $-14.8 $ &  2.9 &  26.1 &  0.844 \\
DF41& $ -1.9 \pm  14.3$ & $ -2.1 \pm  15.7$ & $-14.9 $ &  3.4 &  26.3 &  0.998 \\
DF46& $  0.0 \pm  15.7$ & $  0.0 \pm  22.7$ & $-14.6 $ &  3.4 &  26.6 &  1.206 \\
DF47& $  2.4 \pm  16.9$ & $  2.9 \pm  20.3$ & $-14.8 $ &  4.2 &  26.9 &  1.630 \\
Y112$^b$& $  2.1 \pm   5.6$ & $  4.5 \pm  17.6$ & $-14.2 $ &  1.8 &  25.7 &  0.259 \\
Y121& $ 22.4 \pm   4.1$ & $ 56.3 \pm  30.9$ & $-14.0 $ &  1.5 &  25.5 &  0.295 \\
Y122& $-12.8 \pm   8.1$ & $-38.7 \pm  36.9$ & $-13.8 $ &  2.4 &  26.7 &  0.301 \\
Y358$^b$& $ 28.0 \pm   5.3$ & $ 33.7 \pm  19.1$ & $-14.8 $ &  2.3 &  25.6 &  0.111 \\
Y367$^b$& $ 22.4 \pm   5.2$ & $ 74.2 \pm  44.2$ & $-13.7 $ &  1.7 &  26.0 &  0.093 \\
Y370& $ 17.0 \pm   6.4$ & $ 46.8 \pm  38.6$ & $-13.9 $ &  2.1 &  26.3 &  0.068 \\
Y386& $  6.8 \pm  11.4$ & $  9.0 \pm  24.9$ & $-14.7 $ &  2.6 &  26.0 &  0.056 \\
Y424$^c$& $ 16.7 \pm   5.7$ & $348.3 \pm 262.6$ & $-11.7 $ &  1.7 &  28.0 &  0.163 \\
Y425& $ 10.4 \pm   6.8$ & $ 49.7 \pm  61.3$ & $-13.3 $ &  1.8 &  26.6 &  0.152 \\
Y436$^b$& $ 42.4 \pm   7.3$ & $168.8 \pm  67.0$ & $-13.5 $ &  1.7 &  26.2 &  0.199 \\
Y534$^b$& $ 23.2 \pm   5.0$ & $ 63.9 \pm  36.1$ & $-13.9 $ &  1.9 &  26.1 &  0.713 \\

\tableline
\end{tabular}
\end{center}
\tablenotetext{a}{The background UDG at 142 Mpc}
\tablenotetext{b}{Candidate nucleated UDG.}
\tablenotetext{c}{UDG excluded from our analysis due its much lower luminosity compared with other UDGs.}
\end{table*}
\clearpage


\startlongtable
\begin{deluxetable*}{lccccc}
\tablecaption{The globular cluster system properties of Coma dwarf galaxies$^a$\label{tbl:dwarf}}
\tablehead{
\colhead{Name} & \colhead{$N_{\rm GC}$} & \colhead{$S_N$} & \colhead{$M_V^b$} & \colhead{$R_{\rm e,gal}$} & \colhead{$\langle\mu_V\rangle_e$} \\
\colhead{} & \colhead{} & \colhead{} & \colhead{[mag]} & \colhead{[kpc]} & \colhead{[mag/arcsec$^2$] }
}
\startdata
       COMAi13052.942p28435.86& $ -2.1 \pm   7.3$ & $-11.2 \pm  34.5$ & $-13.2 $ &  1.6 &  24.8 \\
       COMAi13045.913p28335.50& $  7.8 \pm   5.9$ & $ 44.6 \pm  31.0$ & $-13.1 $ &  1.6 &  25.0 \\
      SDSSJ130032.61--280331.4& $ 42.3 \pm  13.6$ & $ 11.7 \pm   4.3$ & $-16.4 $ &  4.0 &  23.6 \\
      SDSSJ130027.57--280323.9& $ -1.8 \pm   8.8$ & $ -0.7 \pm   4.5$ & $-16.1 $ &  4.2 &  24.0 \\
      SDSSJ130026.88--280450.7& $ -8.5 \pm   8.0$ & $-14.8 \pm  16.4$ & $-14.4 $ &  2.5 &  24.6 \\
      SDSSJ130037.05--280544.7& $ -6.4 \pm  11.6$ & $-14.7 \pm  32.9$ & $-14.1 $ &  4.2 &  26.0 \\
      SDSSJ130020.39--280413.9& $ -8.6 \pm   9.1$ & $ -4.5 \pm   3.3$ & $-15.7 $ &  1.9 &  22.7 \\
    SDSSJ130020.39--280413.9$^c$& $  5.8 \pm  10.4$ & $  5.3 \pm   7.2$ & $-15.1 $ &  2.9 &  24.2 \\
      SDSSJ130019.08--280508.9& $ -2.1 \pm   8.6$ & $ -3.7 \pm   6.8$ & $-14.4 $ &  1.9 &  24.0 \\
      SDSSJ130021.38--280327.3& $ -3.0 \pm  12.4$ & $ -6.8 \pm  22.8$ & $-14.1 $ &  2.9 &  25.2 \\
      SDSSJ130020.27--280453.2& $  2.1 \pm   9.7$ & $  4.5 \pm   7.4$ & $-14.2 $ &  1.7 &  24.0 \\
      SDSSJ130022.01--280220.9& $  4.3 \pm 149.7$ & $  9.8 \pm  17.7$ & $-14.1 $ &  1.8 &  24.2 \\
      SDSSJ130031.97--280125.1& $ -2.1 \pm 148.9$ & $ -7.7 \pm  31.7$ & $-13.6 $ &  2.2 &  25.1 \\
      SDSSJ130030.05--280134.8& $  4.3 \pm   6.1$ & $ 14.1 \pm  17.3$ & $-13.7 $ &  1.5 &  24.2 \\
      SDSSJ130005.76--280212.1& $ -0.7 \pm   6.3$ & $ -0.9 \pm   8.3$ & $-14.8 $ &  1.7 &  23.4 \\
      SDSSJ130015.68--280146.3& $ 10.7 \pm   8.9$ & $ 22.3 \pm  20.4$ & $-14.2 $ &  2.5 &  24.8 \\
      COMAi125937.975p28054.48& $ -2.2 \pm   7.6$ & $-12.4 \pm  48.5$ & $-13.1 $ &  2.0 &  25.5 \\
      COMAi125925.477p28211.03& $ -5.6 \pm   7.1$ & $-15.5 \pm  20.0$ & $-13.9 $ &  1.9 &  24.5 \\
      SDSSJ125905.94--280228.8& $ 14.9 \pm  15.2$ & $  6.5 \pm   6.0$ & $-15.9 $ &  3.6 &  23.9 \\
      SDSSJ130034.32--275817.6& $  2.1 \pm   7.6$ & $  4.5 \pm  15.5$ & $-14.2 $ &  1.7 &  23.9 \\
      SDSSJ130039.32--275748.4& $  5.4 \pm   8.3$ & $ 11.3 \pm  18.4$ & $-14.2 $ &  2.1 &  24.5 \\
      COMAi13043.723p275920.89& $-10.7 \pm   6.3$ & $-42.4 \pm  24.3$ & $-13.5 $ &  1.7 &  24.7 \\
    SDSSJ130022.65--275754.9$^c$& $  3.2 \pm  11.1$ & $  0.9 \pm   4.4$ & $-16.4 $ &  2.8 &  22.9 \\
      COMAi13026.767p275953.57& $  4.1 \pm   6.3$ & $  9.4 \pm  14.6$ & $-14.1 $ &  1.6 &  23.9 \\
      COMAi13029.716p275806.72& $  2.1 \pm   5.7$ & $  9.3 \pm  25.2$ & $-13.4 $ &  1.4 &  24.3 \\
      COMAi13023.227p275948.84& $  8.5 \pm   9.3$ & $ 34.0 \pm  43.0$ & $-13.5 $ &  2.0 &  25.0 \\
      SDSSJ125955.93--275748.6& $ 10.7 \pm  14.9$ & $  7.4 \pm  10.2$ & $-15.4 $ &  2.2 &  23.3 \\
      SDSSJ125952.18--275946.3& $  2.1 \pm  11.1$ & $  5.4 \pm  25.0$ & $-14.0 $ &  1.7 &  24.2 \\
     COMAi125956.527p275909.33& $ 10.7 \pm  11.5$ & $ 38.7 \pm  42.7$ & $-13.6 $ &  1.9 &  24.8 \\
      SDSSJ125934.39--275942.9& $  6.4 \pm  11.1$ & $  8.4 \pm  31.8$ & $-14.7 $ &  2.1 &  23.9 \\
      SDSSJ130032.96--275406.6& $ 29.9 \pm  14.9$ & $ 10.9 \pm   6.3$ & $-16.1 $ &  5.8 &  24.7 \\
      COMAi13033.689p275524.77& $  4.3 \pm   6.1$ & $ 22.4 \pm  27.5$ & $-13.2 $ &  1.7 &  25.0 \\
      SDSSJ130022.90--275515.1& $ -4.3 \pm   6.9$ & $ -3.9 \pm   5.3$ & $-15.1 $ &  2.4 &  23.8 \\
      SDSSJ130016.68--275638.9& $  0.0 \pm  15.3$ & $  0.0 \pm   8.6$ & $-15.7 $ &  4.3 &  24.5 \\
      COMAi13025.049p275637.93& $  8.6 \pm   7.8$ & $ 13.6 \pm   9.2$ & $-14.5 $ &  1.5 &  23.5 \\
      SDSSJ130016.37--275522.2& $ 14.9 \pm   7.5$ & $ 14.9 \pm   8.1$ & $-15.0 $ &  2.3 &  23.9 \\
      COMAi13018.409p275530.52& $  4.3 \pm   6.9$ & $ 26.9 \pm  31.8$ & $-13.0 $ &  1.6 &  25.0 \\
      COMAi13021.712p275650.16& $ 12.8 \pm   5.7$ & $ 67.0 \pm  31.3$ & $-13.2 $ &  1.7 &  25.0 \\
      SDSSJ130011.41--275436.4& $  2.0 \pm   4.9$ & $  0.6 \pm   2.2$ & $-16.3 $ &  2.8 &  23.0 \\
      SDSSJ130005.34--275628.9& $  5.3 \pm  10.8$ & $  2.3 \pm   2.9$ & $-15.9 $ &  1.7 &  22.3 \\
      SDSSJ125958.22--275410.8& $ 14.7 \pm  10.0$ & $  7.7 \pm   5.1$ & $-15.7 $ &  3.0 &  23.7 \\
      SDSSJ130010.38--275617.0& $  0.0 \pm  10.0$ & $  0.0 \pm   3.0$ & $-14.3 $ &  0.8 &  22.1 \\
      SDSSJ130007.04--275416.8& $  2.1 \pm   5.1$ & $  5.4 \pm  13.6$ & $-14.0 $ &  1.8 &  24.3 \\
     COMAi125944.825p275536.87& $  8.5 \pm  18.1$ & $ 14.8 \pm  33.0$ & $-14.4 $ &  3.8 &  25.5 \\
     COMAi125956.755p275615.80& $  6.4 \pm  10.2$ & $ 16.1 \pm  25.5$ & $-14.0 $ &  2.2 &  24.7 \\
      SDSSJ125927.22--275257.0& $  4.3 \pm   7.5$ & $  4.3 \pm   7.6$ & $-15.0 $ &  2.1 &  23.6 \\
     COMAi125828.358p271315.01& $ 10.7 \pm   7.6$ & $  4.2 \pm   2.6$ & $-16.0 $ &  3.4 &  23.7 \\
      SDSSJ125636.63--271503.6& $  0.0 \pm  10.7$ & $  0.0 \pm  10.6$ & $-15.1 $ &  2.8 &  24.2 \\
      SDSSJ125844.37--274740.9& $ -6.4 \pm  18.8$ & $ -2.3 \pm   2.4$ & $-16.1 $ &  2.7 &  23.1 \\
	  SDSSJ125856.95--274719.9& $  2.9 \pm  13.2$ & $  4.2 \pm  13.5$ & $-14.6 $ &  2.1 &  24.0 \\
      SDSSJ125843.28--274721.1& $  0.0 \pm  10.2$ & $  0.0 \pm  15.0$ & $-13.8 $ &  1.5 &  24.2 \\
     COMAi125713.240p272437.24& $  4.6 \pm   9.9$ & $  1.5 \pm   2.1$ & $-16.2 $ &  2.1 &  22.5 \\
      SDSSJ125712.27--272313.0& $ 23.5 \pm  11.6$ & $ 10.2 \pm   5.0$ & $-15.9 $ &  3.9 &  24.1 \\
      SDSSJ125711.01--273142.2& $ -4.3 \pm   7.6$ & $ -1.7 \pm   2.3$ & $-16.0 $ &  2.4 &  22.9 \\   
\enddata
\tablenotetext{a}{Coordinates of objects are presented in the Table A1 of \citet{denB14}. }
\tablenotetext{b}{V-band magnitudes are converted from their I-band magnitude with an assumption of dwarf galaxy's color with $V-I=0.4$~mag.}
\tablenotetext{c}{There are duplicated objects with same name. This is the second among them.}
\end{deluxetable*}

\end{document}